\pdfoutput=1 

\documentclass{elsart}

\usepackage{amssymb}
\usepackage{xspace}

\usepackage{listings}

\RequirePackage{ifpdf}
\ifpdf
  \usepackage[pdftex]{graphicx}
  \usepackage[pdftex,colorlinks]{hyperref}
\else
  \usepackage[dvips]{graphicx}
  \usepackage[dvips]{hyperref}

\fi

\usepackage{lineno}

\def\lhcb {LHC{\em b\/}\xspace}
\def\dirac{DIRAC\xspace}
\def\atlas {ATLAS\xspace}
\def\etal {\textit{et al.}}
\def\lhc {LHC\xspace}
\def\ganga {\textsc{Ganga}\xspace}
\def\python {\textsc{Python}\xspace}
\def\root {\textsc{Root}\xspace}
\def\gaudi {\textsc{Gaudi}\xspace}
\def\athena {\textsc{Athena}\xspace}
\def\garfield {\textsc{Garfield}\xspace}
\def\diane {\textsc{DIANE}\xspace}
\def\grid {Grid\xspace}
\def\GPI{GPI\xspace}
\def\roofit{\textsc{RooFit}\xspace}

\def\ARC{ARC\xspace}
\newcommand{\code}[1]{\texttt{#1}}
\newcommand{\val}[1]{\emph{#1}}

\newcommand{\qq}[1]{#1}
\newcommand{\CPCProgramSummary}[1]{}

\begin{document}

\begin{frontmatter}


\title{Ganga: a tool for computational-task \\management and easy access\\ to \grid resources}
\author[a:CERN]{J.T.~Mo{\'s}cicki\corauthref{cor1}},
\corauth[cor1]{Corresponding author}
\ead{jakub.moscicki@cern.ch}
\author[a:Cambridge]{F.~Brochu},
\author[a:Munich]{J.~Ebke}
\author[a:Imperial]{U.~Egede},
\author[a:Munich]{J.~Elmsheuser},
\author[a:Cambridge]{K.~Harrison},
\author[a:Lancaster]{R.W.L.~Jones},
\author[a:NIKHEF]{H.C.~Lee,\thanksref{HurngChun}},
\author[a:CERN]{D.~Liko},
\author[a:CERN]{A.~Maier},
\author[a:CERN]{A.~Muraru},
\author[a:STFC]{G.N.~Patrick},
\author[a:Oslo]{K.~Pajchel},
\author[a:Imperial]{W.~Reece},
\author[a:Oslo]{B.H.~Samset},
\author[a:Birmingham]{M.W.~Slater},
\author[a:Oxford]{A.~Soroko},
\author[a:Birmingham]{C.L.~Tan},
\author[a:CERN]{D.C.~Vanderster}
\author[a:Imperial]{M.~Williams}

\address[a:Cambridge]{University of Cambridge, Cambridge, United Kingdom}
\address[a:Imperial]{Imperial College London, London, United Kingdom}
\address[a:Munich]{Ludwig-Maximilians-Universit\"{a}t, Munich, Germany}
\address[a:Lancaster]{Lancaster University, Lancaster, United Kingdom}
\address[a:NIKHEF]{NIKHEF, Amsterdam, The Netherlands}
\address[a:CERN]{CERN, Geneva, Switzerland}
\address[a:STFC]{STFC Rutherford Appleton Laboratory, Didcot, United Kingdom}
\address[a:Oxford]{University of Oxford, Oxford, United Kingdom}
\address[a:Birmingham]{University of Birmingham, Birmingham, United Kingdom}
\address[a:Oslo]{University of Oslo, Oslo, Norway}

\thanks[HurngChun]{On leave from University of Insbruck, Austria.}

\begin{abstract}

In this paper, we present the computational task-management
tool \ganga, which allows for the specification, submission,
bookkeeping and post-processing of computational tasks on a wide set
of distributed resources.  \ganga has been developed to solve a
problem increasingly common in scientific projects, which is that
researchers must regularly switch between different processing
systems, each with its own command set, to complete their
computational tasks. \ganga provides a homogeneous environment for
processing data on heterogeneous resources. We give examples from High
Energy Physics, demonstrating how an analysis can be developed on a
local system and then transparently moved to a \grid system for
processing of all available data.  \ganga has an API that can be used
via an interactive interface, in scripts, or through a GUI. Specific
knowledge about types of tasks or computational resources is provided
at run-time through a plugin system, making new developments easy to
integrate. We give an overview of the \ganga architecture, give
examples of current use, and demonstrate how \ganga can be used in
many different areas of science.
\end{abstract}

\end{frontmatter}

\begin{linenumbers}

\section{Introduction}
\label{sec:intro}

\qq{
Scientific communities are using a growing number of distributed
systems, from local batch systems and community-specific services to
generic, global Grid infrastructures.  Users may debug applications
using a desktop computer, then perform small-scale application testing
using local resources and finally run at full-scale using globally
distributed Grids. Sometimes new resources are made available to the
users through systems previously unknown to them, and signficant
effort may be required to gain familiarity with these systems
interfaces and idiosyncracies . The time cost of mastering application configuration,
tracking of computational tasks, archival and access to the results is
prohibitive for the end-users if they are not supported by appropriate
tools.  }

\ganga is an easy-to-use frontend for the configuration, execution, and
management of computational tasks. The implementation uses an object-oriented
design in \python~\cite{python}. It started as a project
to serve as a \grid user interface for data
analysis within the \atlas~\cite{ATLAS} and
\lhcb~\cite{LHCb} experiments in High Energy Physics where large
communities of physicists need access to \grid resources for data
mining and simulation tasks. A list of projects which supported the
development of \ganga may be found in
section \ref{sec:acknowledgements}.

\ganga provides a simple but flexible programming interface that can
be used either interactively at the \python prompt, through a
Graphical User Interface~(GUI) or programmatically in scripts. The
concept of a \emph{job} component is essential as it contains the full
description of a computational task, including: the code to execute;
input data for processing; data produced by the application; the
specification of the required processing environment; post-processing
tasks; and metadata for bookkeeping.  The purpose of \ganga can then
be seen as making it easy for a user to create, submit and monitor the
progress of jobs. \ganga keeps track of all jobs and their status
through a repository that archives all information between independent
\ganga sessions. It is possible to switch between executing a job on a
local computer and executing on the \grid by changing a single parameter of a job object. 
This simplifies the progression from rapid prototyping on a local
computer, to small-scale tests on a local batch system, to the analysis of a
large dataset using \grid resources.

In \ganga, the user has programmatic access through an Application
Programming Interface (API), and has access to applications locally for
quick turnaround during development. 

\qq{ \ganga is a user- and application-oriented layer above 
existing job submission and management technologies, such as
Globus \cite{Globus}, Condor \cite{Condor}, Unicore \cite{Unicore} or
gLite \cite{andreetto_2008}. Rather than replacing the existing
technologies, \ganga allows them to be used interchangeably, using a
common interface as the interoperability layer. }

It is possible to make \ganga available to a user community with a high level
of customisation. For example, an expert within a field can implement a custom
application class describing the specific computational task. The class will
encapsulate all low-level setup of the application, which is always
the same, and only expose a few parameters for configuration of a particular
task. The plugin system provided in \ganga means that this expert
customisation will be integrated seamlessly with the core of \ganga at runtime,
and can be used by an end-user to process tasks in a way that requires
little knowledge about the interfaces of \grid or batch systems. Issues such as
differences in data access between jobs executing locally and on the
\grid are similarly hidden.

\ganga may be used as a job management system integrated into a larger system. In this
case \ganga acts as a library for job submission and control. In particular,
\ganga may be used as a building block for the implementation of \grid Portals
which allow users access to \grid functionality through
their web browsers in a simplified way. These portals are normally
domain specific and allow users of a distributed application to run it
on the \grid without needing to know much about \grid tools.

\ganga is licensed under the GNU General Public
License\footnote{\ganga is licensed under GPL version 2 or, if preferred by
the user, any later version.  Details of the GPL are available at
\url{http://www.gnu.org/licenses/gpl.html}.} and is available for download from
the project website: \url{http://www.cern.ch/ganga}. The installation
of \ganga is trivial and does not require privileged access or any
server configuration. \qq{The \ganga installer script provides a
self-contained package and most of the external dependencies are resolved
automatically. However, \ganga generally does not attempt to
install \grid or batch submission tools or the application
software\footnote{Some external dependencies, such as NorduGrid submission tools, are automatically installed.}. Typically such software is installed and managed separately
by system administrators.  Simple configuration files allow 
customisation and configuration of \ganga at the level of site, workgroup and user.
}

Between January 2007 and December 2008 \ganga was
used at 150 sites around the world, with 2000 unique users running about 250k 
\ganga sessions\footnote{The usage information was collected from a voluntary
  usage reporting system implemented in \ganga.}.

In this paper, we describe in section~\ref{sec:functionality} the
overall functionality, in section~\ref{sec:implementation} details of
the implementation, and in section~\ref{sec:mon} how the progress of
jobs is monitored. Section~\ref{sec:GUI} gives an overview of the
Graphical User Interface.  In sections~\ref{sec:useHEP}
and~\ref{sec:other} we discuss how \ganga is customised for specific
user communities. Interfacing and embedding \ganga in other frameworks
is presented in section~\ref{sec:GangaInOtherFrameworks}. In
appendix~\ref{sec:examples} we provide some examples of how the API in
\ganga can be used.

\section{Functionality}
\label{sec:functionality}
\begin{figure}[htbp]
  \centering
  \includegraphics[width=14cm]{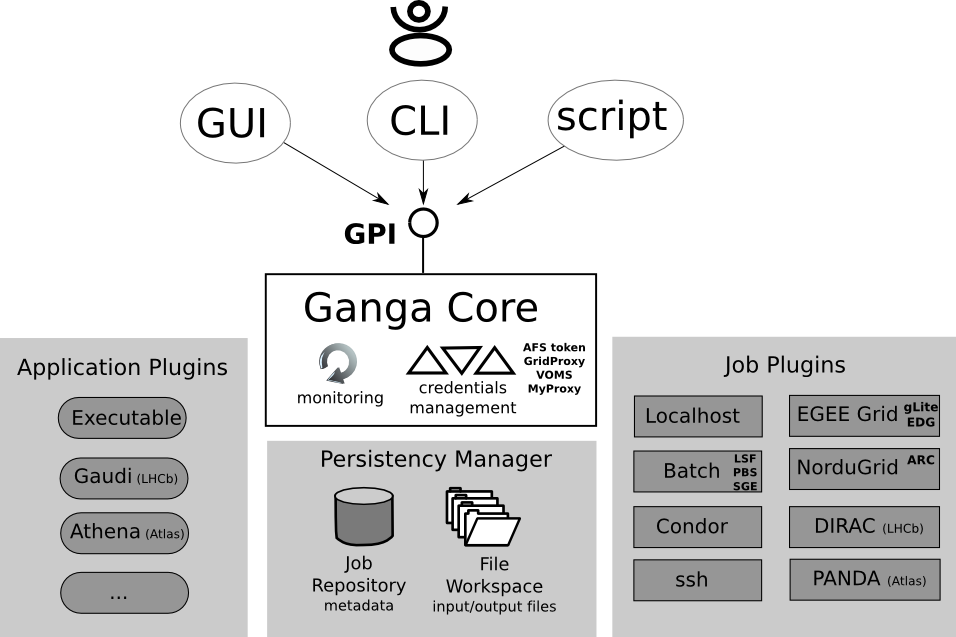}
  \caption{The overall architecture of \ganga. The user interacts with the
    \ganga Public Interface (GPI) via the Graphical User Interface (GUI), the
    Command-Line Interface in Python (CLIP), or
    scripts. Plugins are provided for different application types and
    backends. All jobs are stored in the
    repository.}
  \label{fig:GPI_architecture}
\end{figure}

\ganga is a user-centric tool that allows easy interaction with heterogeneous
computational environments, configuration of the applications and coherent
organisation of jobs. \ganga functionality may be accessed by a user through
any of several interfaces: a text-based command line in \python, a file-based
scripting interface and a graphical user interface~(GUI). This reflects the different
working styles in different user communities, and addresses various usage
scenarios such as using the GUI for training new users, the command line to
exploit advanced use-cases, and scripting for automation of repetitive tasks.
For \ganga sessions the current usage fractions are 55\%, 40\% and 5\%
respectively for interactive prompt, scripts and GUI. As shown in
Fig.~\ref{fig:GPI_architecture}, the three user interfaces are built on top of
the \ganga Public Interface~(\GPI) which in turn provides access to the \ganga
core implementation.

\begin{figure}
  \centering
  \includegraphics[width=14cm]{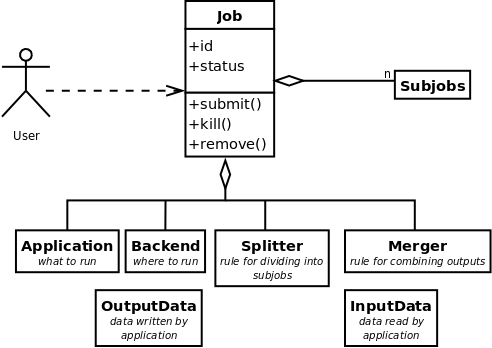}
  \caption{A set of components in \ganga can be combined to form a complete
    job. The application to run and the backend where it will run are
    mandatory while all other components are optional.}
  \label{fig:JobComponents}
\end{figure}

A job in \ganga is constructed from a set of components. All jobs are
required to have an application component and a backend component, which
define respectively the software to be run and the processing system to be
used.  Many jobs also have input and output dataset components,
specifying data to be read and produced.  Finally, computationally intensive
jobs may have a splitter component, which provides a mechanism for dividing
into independent subjobs, and a merger component, which allows for the
aggregation of subjob outputs. The overall component structure of a job is
illustrated in Fig.~\ref{fig:JobComponents}.

By default, the \GPI exposes a simplified, top-level view suitable for most
users in their everyday work, but at the same time allows for the details of
underlying systems to be exposed if needed. An example interactive \ganga
session using the GPI is given in Appendix~\ref{sec:examples}.

\ganga prevents modification by the user of a submitted job.  However,
a copy of the job may easily be created and the copy can be modified.
\ganga monitors the evolution of submitted jobs and categorises
them into the simplified states \val{submitted}, \val{running},
\val{completed}, \val{failed} or \val{killed}.

All job objects are stored in a job repository database, and the input
and output files associated with the jobs are stored in a file workspace. Both
the repository and the workspace may be in a local filesystem or on a remote
server.

A large computational task may be split into a number of subjobs
automatically according to user-defined criteria and the output merged
at a later stage. Each subjob will execute on its own and the merging
of the output will take place when all have finalised. The submission
of subjobs is automatically optimised if the backend component
supports bulk job submission. For example, when submitting to the
gLite workload management system~\cite{andreetto_2008} the job
collection mechanism is used transparently to the user. \qq{ 
Job splitting functionality provides a flat list of subjobs suitable for
parallel processing of fully independent workloads.  However, certain backends
allow users to make use of more-sophisticated parallelisation schemes,
for example the Message Passing Interface (MPI) [8].  In this case, Ganga may be
used to manage collections of subjobs corresponding to MPI processes.}

The \GPI allows frequently used job configurations to be
stored as \code{templates}, so that they may easily be reused, and allows
jobs to be labelled and organised in a hierarchical \code{jobtree}.

\ganga has built-in support for handling user credentials, including
classic \grid proxies, proxies with extensions for a Virtual Organisation Management
Service (VOMS)~\cite{VOMS}, and Kerberos~\cite{kerberos} tokens
for access to an Andrew filesystem (AFS)~\cite{AFS}. A user may renew and destroy the
credentials directly using the GPI. \ganga gives an early warning to a
user if the credentials are about to expire. The minimum credential
validity and other aspects of the credential management are fully
configurable.

\qq{
\ganga supports multiple security models.
For local and batch backends, the authentication and authorisation of
the users is based on the local security infrastructure including user
name and network authentication protocols such as Kerberos. \grid
security infrastructure (GSI) \cite{GSI} provides for security across
organizational boundaries for the \grid backends.  Different security
models are encapsulated in pluggable components, which may be
simultaneously used in the same \ganga session.  }

A \code{Robot} has been implemented for repetitive use-cases. It is a \GPI
script that periodically executes a series of
actions in the context of a \ganga session.  These actions are defined by
implementations of an action interface.  Without programming, the driver can be
configured using existing action implementations to submit saved jobs, wait
for the jobs to complete, extract data about the jobs to an XML file, generate
plain text or HTML summary reports, and email the reports to interested
parties. Custom actions can easily be added by either extending or aggregating
the existing implementations or implementing the action interface directly,
allowing for a diverse variety of repetitive use-cases. An example is given
in section~\ref{sec:lhcb}.

Details of the different kinds of \ganga component are given below, along with
generic examples. More specialised components, designed for a particular
problem domain, are considered in sections \ref{sec:useHEP} and \ref{sec:other}.

\subsection{Application components}

The application component describes the type of computational task to be
performed.  It allows the characteristics and settings of some
piece of software to be defined, and provides methods specifying
actions to be taken before and after a job is processed.  The
pre-processing (configuration) step typically involves examination of
the application properties, and may derive secondary
information. For example, intermediate configuration files for the
application may be created automatically. The post-processing step can
be useful for validation tasks such as determining the validity
of the application output.

The simplest application component (\texttt{Executable}) has three properties:
\begin{description}
\item[\code{exe :}] the path to an executable binary or script;
\item[\code{args:}] a list of arguments to be passed to the executable;
\item[\code{env :}] a dictionary of environment variables and the values they
  should be assigned before the executable is run.
\end{description}
The configuration method carries out integrity checks -- for example
ensuring that a value has been assigned to the \code{exe} property.

\subsection{Backend components}
A backend component contains parameters describing the
behaviour of a processing system. The list of
parameters can vary significantly from one system to another, but can include,
for example, a queue name, a list of requested sites, the minimum memory
needed and the processing time required. In addition, some parameters hold
information that the system reports back to the user, for example the 
system-specific job identifier and status, and the machine where a
job executed.

A backend component provides methods for submitting jobs, and for cancelling
jobs after submission, when this is needed.  It also provides methods for
updating information on job status, for retrieving output of completed jobs
and for examining files produced while a job is running.

Backend components have been implemented for a range of widely used
processing systems, including: local host, batch systems
(Portable Batch System (PBS)~\cite{henderson_1995},
Load Sharing Facility (LSF)~\cite{schwickerath_2008},
Sun Grid Engine (SGE)~\cite{gentzsch_2001},
and Condor~\cite{thain_2005}), and \grid systems, for example  based on
gLite~\cite{andreetto_2008}, ARC~\cite{ellert_2007} and OSG~\cite{OSG}.
Remote backend component allows jobs to be launched directly on remote machines 
using ssh.  

As an example, the batch backend component defines a
single property that may be set by the user:
\begin{description}
\item[\code{queue~~~~~~:}] name of queue to which job should be submitted, the system default
queue being used if this left unspecified,
\end{description}
and defines three properties for storing system information:
\begin{description}
\item[\code{id~~~~~~~~~:}] job identifier;
\item[\code{status~~~~~:}] status as reported by batch system;
\item[\code{actualqueue:}] name of queue to which job has been submitted.
\end{description}

In addition, a remote-backend component allows a job defined in a \ganga
session running on one machine to be submitted to a processing system
known to a remote machine to which the user has access.  For example,
a user who has accounts on two clusters may submit jobs to the batch system
of each from a single machine.

\subsection{Dataset components}
Dataset components generally define properties that uniquely identify a
particular collection of data, and provide methods for obtaining information
about it, for example its location and size. The details of how data
collections are described can vary significantly from one problem domain to
another, and the only generic dataset component in \ganga represents a null
(empty) dataset.  Other dataset components are specialised for use with a
particular application, and so are discussed later.

A strict distinction is made between the datasets and the sandbox (job)
files.  The former are the files or databases which are stored
externally.  The sandbox consists of files which are transferred from
the user's filesystem together with the job.  The sandbox mechanism is
designed to handle small files (typically up to 10MB) while the
datasets may be arbitrarily large.

\subsection{Splitter components}
Splitter components allow the user to specify the number of subjobs to be
created, and the way in which subjobs differ from one another. As an example,
one splitter component (\code{ArgSplitter}) deals with executing the same task
many times over, but changing the arguments of the application executable each
time. It defines a single property:
\begin{description}
\item[args:] list of sets of arguments to be passed to an application.
\end{description}
Specialised splitters deal with creating subjobs that process different parts
of a dataset.

\subsection{Merger components}
Merger components deal with combining the output of
subjobs. Typical output includes files containing data in a
particular format, for example text strings or data representing
histograms. As examples, one merger component (\code{TextMerger})
concatenates the files of standard output and error returned by a set
of subjobs, and another (\code{RootMerger}) sums histograms produced
in ROOT format~\cite{ROOT}. Merging may be automatically performed in
the background when \ganga retrieves the job output or it may be
controlled manually by the user.

\section{Implementation}
\label{sec:implementation}
In this section we provide details of the actual implementation of some of the
most important parts of \ganga.

\subsection{Components}
\label{sec:ComponentImplementation}
Job components are implemented as plugin classes, imported by \ganga
at start-up if enabled in a user configuration file. This means that
users only see the components relevant to their specific area of
work. \qq{Plugins developed and maintained by the \ganga
team are included in the main \ganga distribution and are upgraded
automatically when a user installs a newer \ganga version. Currently,
the list includes around 15 generic plugins and around 20 plugins
specific to \atlas and \lhcb. Plugins specific to other user
communities need to be installed separately but could easily be
integrated into the main \ganga distribution.  }

\begin{figure}
  \centering
  \includegraphics[width=14cm]{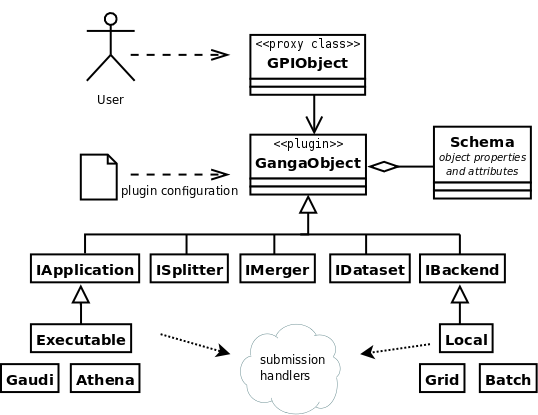}  
  \caption{A component class implements one of the abstract interfaces
    corresponding to the different parts of a job.}
  \label{fig:Components}
\end{figure}
Plugin development is simplified by having a set of internal interfaces and a
mechanism for generating proxy classes~\cite{GoF}. Component classes inherit from an interface class,
as seen in Fig.~\ref{fig:Components}. Each plugin class defines a schema, which
describes the plugin attributes, specifying type
(read-only, read-write, internal), visibility, associated user-convenience filters
and syntax shortcuts.

The user does not interact with the plugin class directly but rather with an
automatically generated proxy class, visible in the \GPI. The proxy
class only includes attributes defined as visible in the schema and methods
selected for export in the plugin class. This separation of the plugin and
proxy levels is very flexible. At the \GPI level, the plugin implementation
details are not visible; all proxy classes follow the same design logic (for
example, copy-by-value); persistence is automatic, session-level locking
is transparent. In this way the low-level, internal API is
separated from the user-level \GPI.

The framework does not force developers to support all combinations of
applications and backends, but only the ones that are meaningful or interesting. To manage
this, the concept of a {\em submission handler} is introduced. The submission
handler is a connector between the application and backend components. At
submission time, it translates the internal representation of the application
into a representation accepted by a specific backend. This strategy allows
integration of inherently different backends and applications without forcing
a lowest-common-denominator interface.

\qq{
Most of the plugins interact with the underlying backends using shell
commands. This down-to-earth approach is particularly useful for
encapsulating the environments of different subsystems and avoiding
environment clashes. In verbose mode, \ganga prints each command
executed so that a user may reproduce the commands externally if
needed. Higher-level abstractions such as JSDL \cite{JSDL},
OGSA-BES \cite{OGSA-BES} or SAGA API \cite{SAGA} are not currently
used, but specific backends that support these standards could readily
be added.
}

\subsection{Job persistence}
\label{sec:persistence}
The \emph{job repository} provides job persistence in a simple database,
so that any subsequent \ganga session has access to all previously
defined jobs. Once a job is defined in a \ganga session it is automatically
saved in the database. The repository provides a bookkeeping system that can
be used to select particular jobs according to job metadata. The metadata
includes such parameters as job name, type of application, type of submission
backend, and job status. It can readily be extended as required.

\ganga supports both a local and a remote repository. In the
case of the former, the database is stored in the local file system,
providing a standalone solution. 
In the case of the latter,
the client accesses an AMGA~\cite{AMGA} metadata
server. The remote server supports secure connections with user
authentication and authorisation based on \grid certificates.
Performance tests of both the local and remote repositories show good
scalability for up to 10 thousand jobs per user, with the average time
of individual job creation being about 0.2 seconds. There is scope for
further optimisation in this area by taking advantage of bulk
operations and job loading on demand.

The job repository also includes a mechanism to support schema migration,
allowing for evolution in the schema of plugin components.

\subsection{Input and output files}

\ganga stores job input and output files in a \emph{job workspace}. 
The current implementation uses the local file system, and has a simple
interface that allows transparent access to job files within the
\ganga framework. These files are stored for each job in a separate
directory, with sub-directories for input and output and for each subjob.

Users may access the job files directly in the file-system or using \ganga commands
such as \texttt{job.peek()}. Internally, \ganga handles the input and output
files using a simple abstraction layer which allows for trivial integration
of additional workspace implementations.
Tests with a prototype using a WebDav~\cite{WebDav}
server have shown that all workspace data related to a
job can be accessed from different locations. In this case, a workspace
cache remains available on the local file system.

The combination of a remote workspace and a remote job repository effectively
creates a roaming profile, where the same \ganga session can be accessed at
multiple locations, similar to the situation for accessing e-mail messages
on an IMAP~\cite{IMAP} server.

\section{Monitoring}
\label{sec:mon}
\ganga provides two types of monitoring: the internal monitoring updates
the user with information on the progress of jobs, and the external
monitoring deals with information from third-party services.

\subsection{Internal monitoring}
\label{sec:GangaMonitoring}
\begin{figure}[htbp]
  \begin{center}
    \includegraphics[width=1 \textwidth]{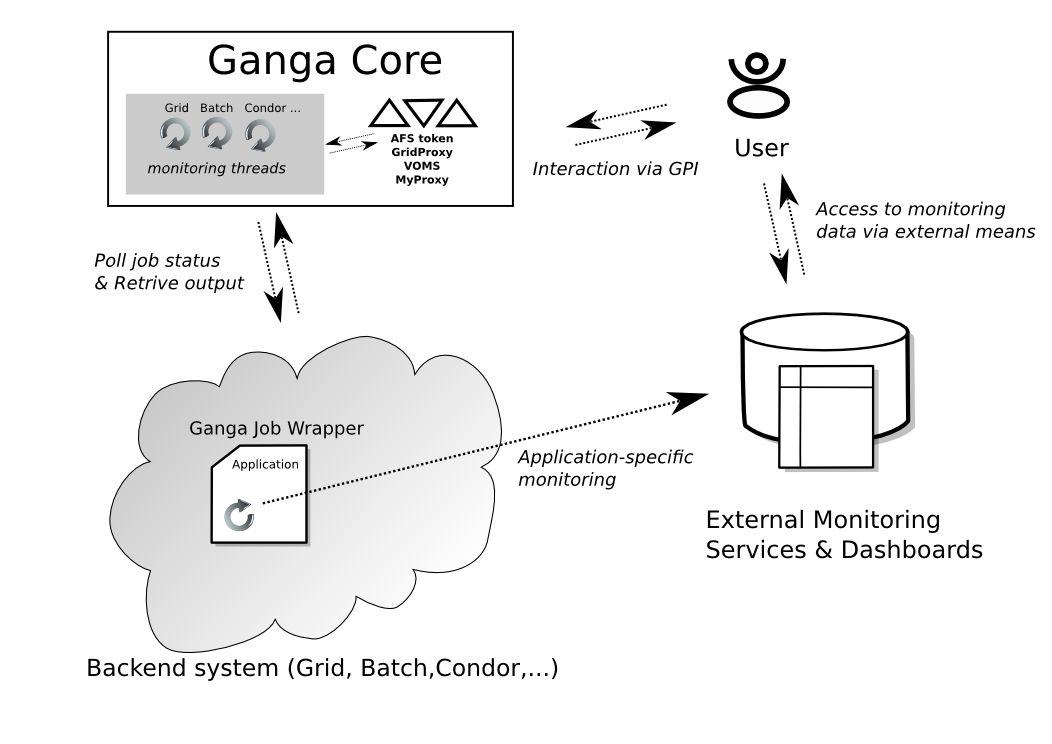}
    \caption{The internal monitoring updates the status of jobs using a pool of threads running in the \ganga core. Additional monitoring thread runs in a job wrapper and sends the monitoring information to external services. }
    \label{fig:job_status_monitoring_mechanism}
  \end{center}
\end{figure}

\ganga automatically keeps track of changes in job status, using a
monitoring procedure designed to cope with
varying backend response times and load capabilities. As seen in
Fig.~\ref{fig:job_status_monitoring_mechanism}, each backend is polled in a
different thread taken from a pool, and there is an efficient mechanism
to avoid deadlocks from backends that respond slowly. The poll rate may be
set separately for each backend.

The monitoring sub-system also keeps track of the remaining validity of
authentication credentials, such as \grid proxies and Kerberos tokens.
The user is notified that renewal is required, and if no action is
taken then \ganga is placed in a state where operations requiring
valid credentials are disabled.

\subsection{External Monitoring}
\label{sec:ExternalMonitoring}
\ganga's external monitoring provides a mechanism for dynamically adding third-party
monitoring sensors, to allow reporting of different metrics for running jobs.

The monitoring sensors can be inserted both on the client side - where \ganga
runs - and on the remote environment (worker node) where the application
runs, allowing the user to follow the entire execution flow.  Monitoring
events are generated at job submission time, at startup, periodically
during execution, and at completion.

Individual application and backend components in \ganga can be configured to use
different monitoring sensors, allowing collection of both generic execution
information and application-specific data.

Use is currently made of two implementations of external monitoring sensors. One
is the \atlas Dashboard application monitoring~\cite{andreeva_2008}.
Another is a custom service that allows the \ganga user to examine job output
in real-time on the \grid.  This streaming service
is not enabled by default, but must be set up for each user community separately,
and may then be requested by a user for specific jobs.

\section{Graphical User Interface}
\label{sec:GUI}

\begin{figure}[ht!]
  \centering
  \includegraphics[width=1 \textwidth]{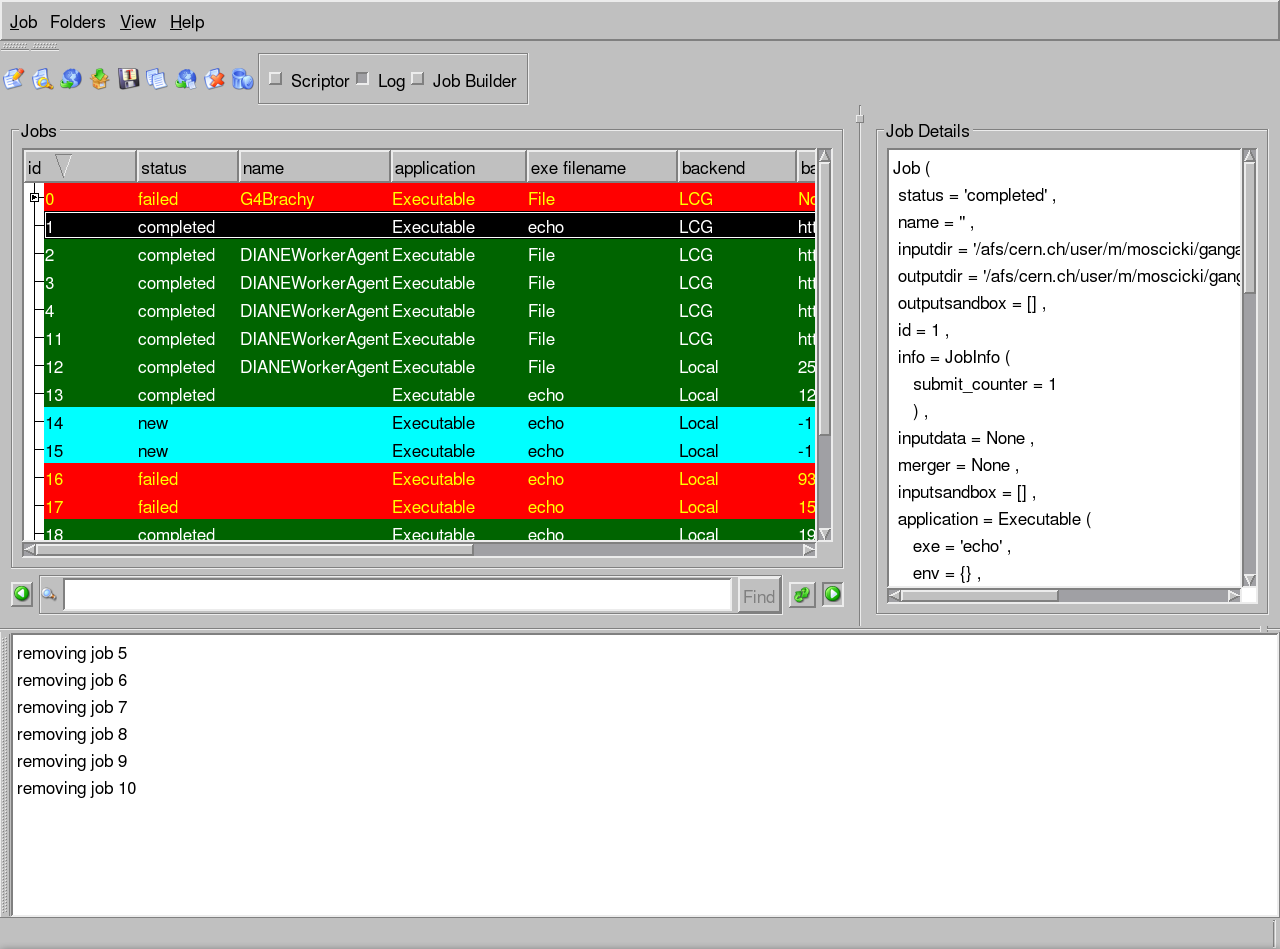}
  \caption{\ganga graphical user interface (GUI).  The overview of jobs
   can be seen to the left, and the details of an individual job are to
   the right.}
  \label{fig:GUI}
\end{figure}

The \ganga Graphical User Interface (GUI), shown in  Fig.~\ref{fig:GUI} and
built using PyQt3~\cite{rempt_2001}, makes available all of the job-management
functionality provided at the level of the \ganga Public Interface.  The
GUI incorporates various convenience features, and its multi-threaded nature
results in a degree of parallelism not possible at the command line:
job monitoring and most job-management actions run concurrently, ensuring
a good response time for the user.

The job monitoring window takes centre stage,
with job status and other monitored attributes displayed
in table format. Other features include subjob monitoring, subjob  
folding/hiding, a job-details display drawer, a logical-collections  
drawer, and a text-based job-search facility. Many characteristics of the
monitoring window can be customised, allowing, for example, selection of
the job attributes to be monitored, and of the colours used to denote
different job states.

The construction of a job, entailing selection of the required plugins and  
the entry of attribute values, is achieved from a job-builder window. This
displays a foldable tree of job attributes, and associated data-entry
widgets. The tree and widgets are generated dynamically based
on plugin schemas, ensuring that the GUI automatically supports
user-defined plugins without any change being needed to the GUI code.  To
assist with data entry, drop-down menus list allowed values, wherever
these are defined; and tool tips provide explanations of individual job
attributes.  The job-builder window also features tool buttons for
performing a wide range of job-related actions, including
creation, saving, copying, submission, termination and removal.  Finally,
a multifunction \code{Extras} tool button provides access to arbitrary
additional functionality implemented in the plugins.

The GUI also has a scriptor window, providing a
favourite-scripts collection, a job-script editor and an embedded \python
session.  The favourite-scripts collection allows frequently used \ganga
scripts to be created, imported, exported and cloned; the job-script editor
facilitates quick modification and execution of scripts; and the
embedded \python session allows interactive use of \ganga commands.

Finally, a scrollable log window collects and displays all messages  
generated by \ganga.

\section{Use in experiments at the Large Hadron Collider}
\label{sec:useHEP}

The \atlas and \lhcb experiments aim to make discoveries about the
fundamental nature of the Universe by detecting new particles at 
high energies, and by performing high-precision measurements of
particle decays. The experiments
are located at the Large Hadron Collider~(\lhc) at the European Laboratory for
Particle Physics~(CERN), Geneva, with first particle collisions (events) expected in
2009. Both experiments require processing of data
volumes of the order of petabytes per year, rely on computing resources
distributed across multiple locations, and exploit several \grid implementations. The data-processing applications,
including simulation, reconstruction and final analysis for the experiments,
are based on the \code{C++} \gaudi/\athena~\cite{gaudi} framework.  This
provides core services, such as message logging, data access, histogramming,
and a run-time configuration system. 

The data from the experiments will be distributed at computing facilities
around the world. Users performing data analysis need an on-demand access
mechanism to allow rapid pre-filtering of data based on certain selection
criteria so as to identify data of specific interest.

The role of \ganga within \atlas and \lhcb is to act as the interface for data
analysis by a large number of individual physicists. \ganga also allows for
the easy exchange of jobs between users, something that can otherwise be difficult
because of the complex configuration of analysis jobs.

\subsection{The \lhcb experiment}
\label{sec:lhcb}

The \lhcb experiment is dedicated to studying the properties of \textit{B}
mesons (particles containing the \textit{b} quark) and in this section we
describe the way in which \ganga interacts with the application and
backend plugins specific to \lhcb.

In a typical analysis, users supply their own shared libraries, containing
user-written classes, and these are loaded at run-time. 
The \lhcb applications are driven by a configuration file,
which includes definitions of the libraries to load, non-default values for
object parameters, the input data to be read, and the output to be created.

\ganga includes an application component for \gaudi-based applications to simplify
the task of performing an analysis. During the configuration stage, and before
job submission, the application component undertakes the following tasks:
\begin{itemize}
\item it locally sets up the environment for the chosen application;
\item it determines the user-owned shared libraries required to
  run the job;
\item it parses the configuration file supplied, including all its dependencies;
\item it uses information obtained from the configuration file to determine
  the input data required and the outputs expected;
\item it registers the inputs and outputs with the submission backend.
\end{itemize}
The user, then, only needs to specify the name and
version of the application to run, and the configuration file to be used.

Code under development by a user may contain bugs that cause
runtime errors during job execution. The transparent switching between
processing systems when using \ganga means that debugging can be
performed locally, with quick response time, before launching a large-scale
analysis on the \grid, where response times tend to be longer.

Some studies in \lhcb, rather than being based on \gaudi, are performed using
the \roofit~\cite{RooFit} framework, most notably studies that make use of
simplified event simulations.   Jobs for these studies require large amounts
of processing power, but do not require
input data and produce only small amounts of output. This makes them
very easy to deploy on the \grid, with support in \ganga provided by a
generic \root~\cite{ROOT} application component.

\qq{
In the \lhcb computing model~\cite{lhcb:2005jj}, \grid jobs are routed
through the
\dirac~\cite{DIRAC} workload management system~(WMS). \dirac is a pilot-based
system where user jobs are queued in the WMS server and the server
submits generic pilot scripts to the Grid.  Each pilot queries the WMS
for a job with resource requirements satisfied by the machine where
the pilot script is running. If a compatible job is available, it is
pulled from the WMS and started.  Otherwise, the pilot terminates and
the WMS sends a new pilot to the \grid.  This system improves the
reliability of the \grid system as seen by the user. \ganga provides
a \dirac backend component that supports submission of jobs to
the \dirac WMS, making use internally of \dirac's \python
API~\cite{DIRACAPI}.  }

A \emph{splitter} component implemented specifically for \lhcb is able to divide
the
analysis of a large dataset into many smaller subjobs. During the splitting,
a file catalogue is queried to ensure that all data associated with an
individual subjob is
available in its entirety at a minimum of one location on the \grid. This gives
significant optimisation, as it avoids subjobs having to copy data across the
network
before an analysis can start.

In total, above 300k user jobs finished successfully in 2008 with a total
CPU consumption of 87 CPU years. The jobs ran at a total of 140 Grid
sites across the globe. The system was responsive to a highly irregular
usage pattern and spikes of several thousand simultaneous jobs were
observed during the year. This usage is expected to rise dramatically after the start of the
\lhcb data taking.

The \code{Robot} in \ganga is used within \lhcb for \emph{end-to-end} testing
of the distributed analysis model. It submits a representative set of
analysis jobs on a daily basis, monitors their progress, and checks 
the results produced. The overall success rate and the time to obtain
the results is recorded and published on the web. The
\code{Robot} monitors this information, producing statistics on the
long-term system performance.

\subsection{The \atlas experiment}
\label{sec:atlas}

\atlas is a general-purpose experiment, designed to allow observation of new
phenomena in high-energy proton-proton collisions.

The distributed analysis model is part of the \atlas computing
model~\cite{bib:atlascompmod} which requires that data are distributed at
various computing sites, and user jobs are sent to the data.

An \atlas analysis job typically consists of a \python
or shell script that configures and runs user algorithms in the \athena
framework~\cite{bib:atlascompmod}, reads and writes event files, and
fills histograms/n-tuples. More-interactive analysis may be performed on
large datasets stored as n-tuples.

There are several scenarios relevant for a user analysis.  Some analyses require
a fast response time and a high level of user interaction, for which the
parallel \root facility PROOF~\cite{ballintijn_2006} is well suited.  Other
analyses require a low level of user interaction, with long response times
acceptable, and in these cases \ganga and \grid processing are ideal.

Analysis jobs can produce large amounts of data, which may initially be
stored at a single \grid site, and may subsequently need to be transferred
to other machines.  This is supported in \atlas by the Distributed Data
Management system DQ2~\cite{bib:atlasdq2}.  This provides
a set of services for moving data between \grid-enabled computing facilities,
and maintains a series of databases that track the data movements.  The
vast amounts of data involved are grouped into datasets, based on various
criteria, for example physics characteristics, to make queries and retrievals
more efficient.

\subsubsection{\atlas \grid infrastructures}

The \atlas experiment employs three \grid infrastructures for user
analysis and for collaboration-wide event simulation and reconstruction. These
are the \grid developed in the context of Enabling  Grids for e-Science
(EGEE, mainly Europe)~\cite{jones_2005}, accessed using gLite
middleware~\cite{andreetto_2008}, the Open Science Grid (OSG, mainly North
America)~\cite{OSG}, accessed using the PanDA system~\cite{maeno_2008}, and
NorduGrid (mainly Nordic countries)~\cite{ellert_2003}, accessed using the
\ARC middleware~\cite{ellert_2007}.  \ganga seamlessly submits jobs to all
three \grid flavours.

\subsubsection{\atlas user analysis}
A typical \atlas user analysis consists of an event-selection algorithm
developed in the Athena framework. Large amounts of data are filtered to
identify events that meet certain selection criteria. The events of interest are
stored in files grouped together as datasets in the DQ2 system.  The \ganga
components for Athena jobs include the following functionality:
\begin{itemize}
\item During job submission, DQ2 is queried for the file content and location
of the dataset to be analysed.  The number of possible \grid sites is then
restricted to the dataset locations.
\item A job can be divided into several subjobs, each processing a given
number of files from the full dataset.
\item In a \grid job, after the \athena
application has completed, the user output 
is stored on the storage element of the site where the job was run, and is
registered in DQ2.
\end{itemize}

In the second half of 2008, more than $4 \times 10^5$ \grid jobs were submitted
through \ganga by \atlas users.  Following a procedure similar to that of
\lhcb, the \ganga \code{Robot} submits test jobs daily to \atlas \grid
sites.  Test results are used to guide users to sites that are performing
well, avoiding job failures on temporarily misconfigured sites.

\subsubsection{\atlas small-scale event simulations}
In addition to data analysis, users sometimes need to simulate event samples
of the order of a few tens of thousands of events. The \emph{AthenaMC}
application component has
been developed to integrate software used in the official \atlas system
for event simulation.  This component consists of a set of \python classes
that together handle input
parameters, input datasets and output datasets for the three
production steps: event generation, detector simulation, and
event  reconstruction. As in the case of user analysis, 
datasets are managed by
the DQ2 system.

\section{Other usage areas}
\label{sec:other}
\ganga offers a flexible and extensible interface that make it useful
beyond the original scope of particle-physics
applications in the \atlas and \lhcb experiments. Here we provide
details of just a few of the other contexts in which \ganga has been
used.

\subsection{Enabling industrial-scale image retrieval}
\label{sec:Imense}
\begin{figure}[htb]
  \begin{center}
    \includegraphics[width=0.85 \textwidth]{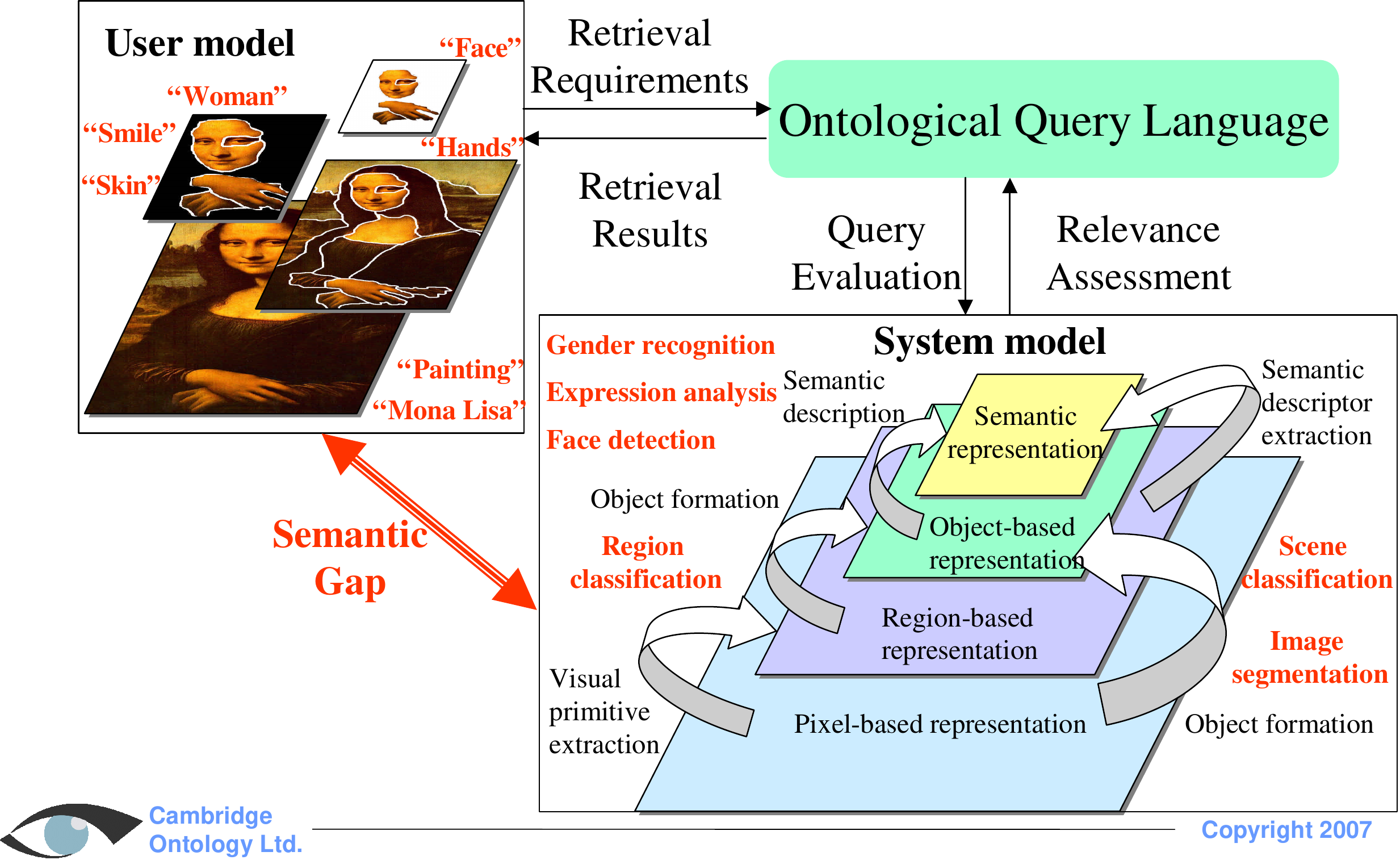}
  \end{center}
  \caption{Schematic representation of the image-retrieval system developed by
Imense Ltd. Image characteristics are determined by applying feature-extraction
algorithms, and an ontological query language bridges the semantic gap between
terms that might be employed in a user query and terms understood by
the processing system.}
  \label{fig:camtologytech}
\end{figure}

Imense Ltd\footnote{\url{http://imense.com}}, a Cambridge-based startup
company, has implemented a novel image retrieval-system
(Fig.~\ref{fig:camtologytech}), featuring automated analysis
and recognition of image content, and an ontological query language. The
proprietary image analysis, developed from published research~\cite{town_2004},
includes recognition of visual properties, such as colour, texture
and shape; recognition of materials, such as grass or sky; detection of
objects, such as human faces, and determination of their characteristics; and
classification of scenes by content, for example beach, forest or
sunset.  The system uses semantic and linguistic relationships between terms to
interpret user queries and retrieve relevant images on the basis of the
analysis results. Moreover, the system is extensible, so that
additional image classification modules or image context and metadata can
easily be integrated into the index.

By using the \ganga framework for job submission and management, it has been
possible to port and deploy a large part of Imense's image-analysis technology
to the \grid and build a searchable index for more than
twenty-million high-resolution photographic images.

The processing stages for the image-search system -- image analysis
and indexing -- are intrinsically sequential.  Analysis has been parallelised
at the level of single
images or small subsets of images. Each image can therefore be processed in
isolation on the \grid, with this processing usually taking a few to ten
seconds.  In order to minimise overheads, images are grouped in sets of
a few hundred per job submitted through \ganga.  Results of the image
processing and analysis are passed back to the submission server once a
job has successfully completed.

Support for Imense has been added to \ganga through the implementation of two
specialised components: an application component that deals with running the
image-processing software, and a dataset component for taking care of
the output. As usual with \ganga, the jobs can run both locally and on the
\grid, giving maximum flexibility.

At runtime, images are retrieved and segmented one at a time, all of the
images are classified, and finally an archive is created of the output files
(several per input image).  The archive is returned using the sandbox
mechanism in \ganga when using the \code{Local} backend, and is uploaded to a
storage element when using the \grid \code{LCG} backend.

The specialised dataset component provides methods for downloading a results
archive from a storage element, and for unpacking an archive to a destination
directory. These methods are invoked automatically by \ganga when an
image-processing job completes: the effect for the user is that a list of
images is submitted for processing and results are placed in the requested
output location independently of the backend used.

\subsection{Smaller collaborations in High Energy Physics}
\label{sec:smallHEP}
Large user communities, such as \atlas and \lhcb, profit from encapsulating
shared use cases as specialised applications in \ganga. In contrast,
individual researchers or developers in the context of rapid prototyping
activities may opt to use generic application components.
In such cases, \ganga still provides the benefits of
bookkeeping and a programmatic interface for job submission. As an example of
this way of working, a small
community of experts in the design of gaseous detectors
use \ganga to run the \garfield~\cite{Garfield} simulation program on the
\grid.  A \ganga script has been written that generates a chain of
simulation jobs
using the \garfield generator of macro files and \ganga's \code{Executable}
application component.  The \garfield executables, and a few small input files,
are placed in
the input sandbox of each job. Histograms and text output are then returned
in the
output sandbox. This simple approach allowed integration of
\garfield jobs in \ganga in just a few hours.

\subsection{\ganga integrated with lightweight \grid middleware}

\qq{
The open-plugin architecture of \ganga allows easy integration of
additional \grid middleware, as has been achieved, for example, with the
\ARC (Advanced Resource Connector) \grid middleware~\cite{ellert_2007}.  This is a
product of the NorduGrid project~\cite{ellert_2003}, and is used by many academic
institutions in the Nordic countries and elsewhere.
}

\ARC jobs are accepted and brokered by a \grid manager, running at site level, and resource
lookup is done through load balancing and runtime environments
advertised by individual sites. File storage and access is 'cloudy',
meaning that all files registered in \grid--wide catalogues are
accessible to all worker nodes. File transfers are handled by
the \grid manager, between job acceptance and
execution. \ARC--connected resources are used e.g. by researchers in
bioinformatics, genomics, meteorology, in addition to high--energy
physics.

\ganga has been interfaced to \ARC through a backend, which translates
\ganga input into \ARC--readable xRSL language. The \ARC user client is
lightweight, and binaries are provided as an external library at \ganga
install time. The main user of this integration is the \atlas
experiment (see sec.~\ref{sec:atlas}), where it is the main user
access portal to one of the experiment's three main computing
grids. Further collaboration between \ARC and \ganga is envisaged, to
employ \ganga as a fully featured frontend to \ARC.

\section{Interfacing to other frameworks}
\label{sec:GangaInOtherFrameworks}
The \ganga Public Interface constitutes an API for generic job submission
and management.  As a result, \ganga may be programmatically
interfaced to other frameworks, and used as a convenient abstraction layer for
job management. \ganga has been used in combination with \diane~\cite{DIANE},
a lightweight agent-based
scheduling layer on top of the \grid, in a number
of scientific activities.  These have included: dosimetry-related simulation
studies in medical physics~\cite{Geant4MedicalDIANE}; regression testing of the Geant~4~\cite{Geant4}
detector-simulation toolkit;
in-silico molecular docking in searches for 
new drugs against potential variants of an influenza virus~\cite{AvianFlu};
telecommunication applications~\cite{ITU}; and theoretical physics~\cite{LQCD}. The
\diane worker agents are executed as \ganga jobs, so that resource usage
may be controlled by the user from the \ganga interface. This
approach allows the efficiency of the \diane overlay scheduling system
to be combined with the well-structured job management offered by \ganga, as well as
combining \grid and non-\grid resources under a uniform interface. 
Also, this allows the efficient implementation of low-latency access to \grid resources and improvements to responsiveness when supporting on-demand computing and interactivity \cite{SchedulingForResponsiveGrids}.

\begin{figure}[h!]
  \centering
  \includegraphics[width=1 \textwidth]{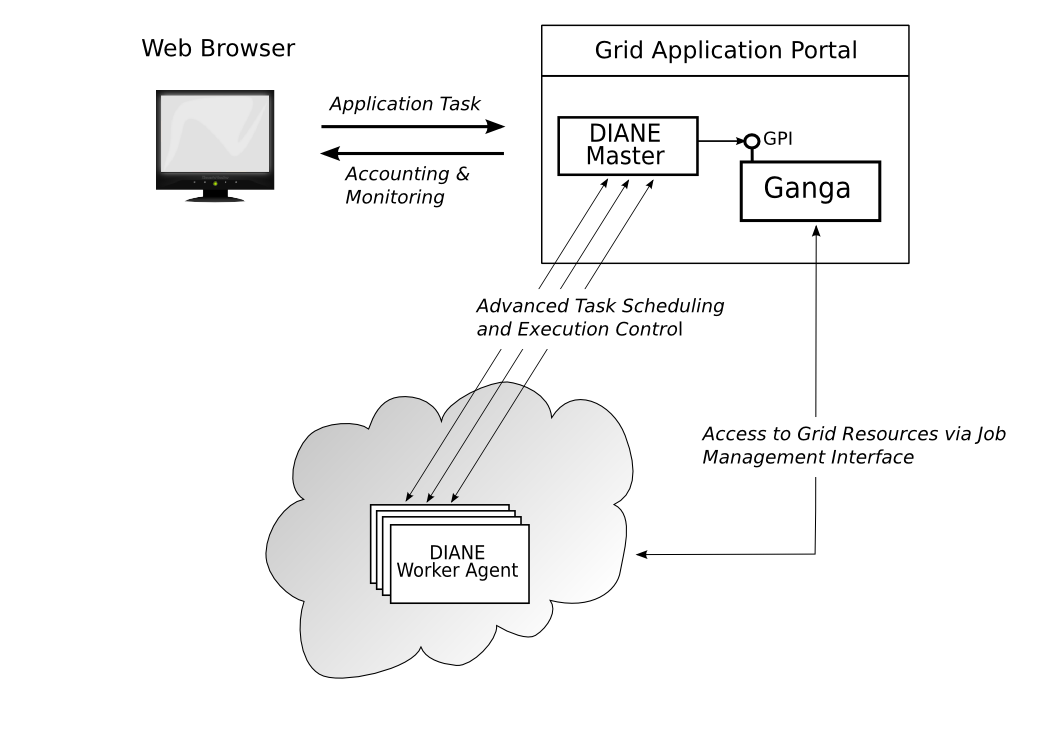}
  \caption{Ganga as a job management component embedded in \diane,
    with an application portal.}
  \label{fig:webportal}
\end{figure}
\ganga may be embedded in web-based services such as the
bio-informatics portal developed by ASGC, Taipei. The portal is fully
customized for analysis of candidate drugs against avian flu.  The portal
engine delegates
job management to the embedded \diane/\ganga framework, as shown in
Fig.~\ref{fig:webportal}. Following this approach, users can
switch between different resources, or access heterogeneous computing environments
through a single same web interface.

\section{Conclusion}
\label{sec:conclusion}
\ganga has been presented as a tool for job management in an environment of heterogeneous resources
and is particularly suited to the \grid paradigm that has emerged in large-scale distributed computing.
\ganga makes it easy to define a
computational task that can be executed locally for debugging, and
subsequently be run on the \grid, for large scale data mining. We have shown how \ganga
simplifies task specification, takes care of job submission, monitoring and
output retrieval, and provides an intuitive bookkeeping system.

We have demonstrated the advantages of having a well-defined API, which can be
used interactively at the \python prompt, through a GUI or
programmatically in scripts. By virtue of its plugin system, \ganga is readily
extended and customised to meet the requirements of new user communities.
Examples of \ganga usage have been provided in particle physics,
medical physics and image processing.

\qq{
Existing command-line submission interfaces, such as gLite, tend to
include only limited usability features.  Some higher level tools,
for example GridWay\cite{GridWay}, present jobs as if they were Unix processes and
corresponding command line utilities. Interfaces based on Condor
job-submission scripts have also been developed \cite{MCS}. A distinctive
feature of \ganga is that it may easily be adapted to different styles
of working, allowing simultaneous use of three different interfaces.
\ganga also provides a higher level of abstraction than most job-management
tools, and allows a user to focus on solving the domain-specific problems,
rather than changing their way of working each time they switch to a new
processing system.
}

%

\ganga has a large user base and is in active development. \ganga is a tool which may easily
be used to support new scientific or commercial projects on a wide range of distributed
infrastructures.

\section{Acknowledgements}
\label{sec:acknowledgements}
The development of \ganga has been supported by the GridPP project in the
United Kingdom~\cite{Faulkner:2006px}, with funding from the Science and Technology
Facilities Council (STFC) and its predecessor, the Particle Physics and
Astronomy Research Council (PPARC); by the D-Grid project in Germany,
with funding from the Bundesministerium f\"ur Bildung und Forschung
(BMBF); and by the project for Enabling Grids for E-scienceE (EGEE),   
co-funded by the European Commission (contract number INFSO-RI-031688) through the Sixth
Framework Programme.

\ganga has received contributions over the years from a number of
individuals.  Particular thanks are due to David Adams, Marcello
Barisonzi, Mike Kenyon, Wim Lavrijsen, Janusz Martyniak, Pere Mato,
Caitriana Nicholson, Rebecca Ronke, Vladimir Romanovski, David
Tuckett, Ruth Dixon del Tufo, Craig Tull.

The developers would also like to thank the large number of users, from both
within and outside particle physics, for their valuable suggestions for
improving \ganga, and for their help in debugging problems.

\appendix

\section{Examples}
\label{sec:examples}
Below we give a set of examples of working with \ganga. For ease of reading,
\python keywords are in bold. First we look at a complete \ganga session.
\vspace{-2ex}

\tiny
\lstset{language=Python} \lstset{commentstyle=\textit}
\lstset{backgroundcolor=,rulecolor=}
\begin{lstlisting}[escapechar=!]{}
!
\begin{verbatim}
~ % ganga
*** Welcome to Ganga ***
Version: Ganga-5-1-0
Documentation and support: http://cern.ch/ganga
Type help() or help('index') for online help.

This is free software (GPL), and you are welcome to redistribute
it under certain conditions; type license() for details.
\end{verbatim}!
[1]: j=Job(name='MyJob')      # Create a default job
[2]: j.submit()               # Submit the job

# wait for the monitoring

[3]: j.peek('stdout')         # Look at the output
[4]: j=j.copy(name='GridJob') # Make a copy of the job
[5]: j.backend=LCG()          # Change backend to the Grid
[6]: j.submit()               # Submit the job
[7]: jobs                     # List jobs
!
\begin{verbatim}
...job listing...
\end{verbatim}!
[8]: Exit                     # Quit Ganga.
\end{lstlisting}
\normalsize

\vspace{-2ex}
In the next example, we create a job for analysis of \lhcb
data. A splitter is used to divide the analysis between subjobs.
Data are assigned using logical identifiers, and the \dirac WMS ensures
that subjobs are sent to locations where the required data are available.
\tiny
\begin{lstlisting}[escapechar=!]{}
[1]: j=Job(application=DaVinci(),backend=Dirac())
[2]: j.inputdata=LHCbDataset(files=[  # Data to read
...      'LFN:/foo.dst',
...      'LFN:/bar.dst',
...      many more data files])
[3]: j.splitter=DiracSplitter()       # We want subjobs
[4]: j.submit()
!
\begin{verbatim}
Job submission output
\end{verbatim}!
\end{lstlisting}
\normalsize

\vspace{-2ex}
Here, we use the fact that standard \python commands are available at the
\ganga prompt, and print information on subjobs.
\tiny
\begin{lstlisting}[escapechar=!]{}
# Status of jobs and where they ran
[5]: for subjob in j.subjobs: 
...       print subjob.status, subjob.actualCE
!
\begin{verbatim}
42
\end{verbatim}!
# Find backend identifier of all failed jobs
[6]: for j in jobs.select(status='failed'):
...       print j.backend.id
!
\begin{verbatim}
42
\end{verbatim}!
\end{lstlisting}
\normalsize

\vspace{-2ex}

Groups of jobs may be accessed and manipulated using simple methods:

\tiny
\begin{lstlisting}[escapechar=!]{}
[1]: jobs.select(status='failed').resubmit()
[2]: jobs.select(name='testjob').kill()
[3]: newjobs = jobs.select(status='new')
[4]: newjobs.select(name='urgent').submit()
\end{lstlisting}

\end{linenumbers}

\normalsize

\vspace{-2ex}

\begin{thebibliography}{00}

\bibitem{python}
G.~van Rossum and F.L.~Drake~Jr.,
\href{http://www.network-theory.co.uk/python/language/}
{\textit{The Python language reference manual: revised and updated for version
2.5}} (Network Theory Limited, Bristol, 2006).

\bibitem{Faulkner:2006px}
P.J.W.~Faulkner \etal\ [GridPP Collaboration],
\textit{GridPP: development of the UK computing Grid for particle physics},
\href{http://dx.doi.org/10.1088/0954-3899/32/1/N01}
{J.~Phys.\ G: Nucl.\ Part.\ Phys.\ \textbf{32} N1}.

\bibitem{ATLAS}
G.~Aad \etal\ [ATLAS Collaboration],
\textit{The ATLAS Experiment at the CERN Large Hadron Collider},
\href{http://dx.doi.org/10.1088/1748-0221/3/08/S08003}
{JINST {\bf 3} (2008) S08003}.


\bibitem{LHCb}
A.A.~Alves Jr.\ \etal\ [LHCb Collaboration],
\textit{The LHCb Detector at the LHC},
\href{http://dx.doi.org/10.1088/1748-0221/3/08/S08005}
{JINST {\bf 3} (2008) S08005}.

\bibitem{Globus}

I.~Foster, C.~Kesselman and S.~Tuecke,
\textit{The Anatomy of the Grid: Enabling Scalable Virtual Organizations},
{International J. Supercomputer Applications, \textbf{15} (3), 2001.}

\bibitem{Condor}
D.~Thain, T.~Tannenbaum, and M.~Livny, 
\textit{Distributed Computing in Practice: The Condor Experience}
{Concurrency and Computation: Practice and Experience, \textbf{17} (2-4), pp 323-356, 2005.}


\bibitem{Unicore}
A.~Streit \etal,
\textit{UNICORE -- From Project Results to Production Grids}
\href{http://www.unicore.eu}
{Grid Computing: The New Frontiers of High Performance Processing
Advances in Parallel Computing 14, Elsevier, 2005, pp. 357-376}

\bibitem{andreetto_2008} P.~Andreett \etal,
\textit{The gLite workload management system},
\href{http://dx.doi.org/10.1088/1742-6596/119/6/062007}
{J.~Phys.: Conf.\ Ser.\ \textbf{119} (2008) 062007}.

\bibitem{thomas_2005} M.P.~Thomas \etal,
\textit{Grid portal achitectures for scientific applications},
\href{http://dx.doi.org/10.1088/1742-6596/16/1/083}
{J.~Phys.: Conf.\ Ser.\ \textbf{16} (2005) 596}.

\bibitem{li_2006} M.~Li and M.~Baker,
\textit{A review of Grid Portal technology}, pp.~126-156 of:
\href{http://www.springer.com/computer/programming/book/978-1-85233-998-2}
{J.C.~Cunha and O.F.~Rana (Eds.),
\textit{Grid computing: software environments and tools}}
(Springer-Verlag London Ltd, 2006).


\bibitem{MPI} M.~Snir and S.~Otto, \textit{MPI-The Complete Reference: The MPI Core}, 
{MIT Press (1998) ISBN: 0262692155}
\href{http://www.mpi-forum.org}.


\bibitem{VOMS} R.~Alfieri \etal,
\textit{From gridmap-file to VOMS: managing authorization in a Grid
environment},
\href{http://dx.doi.org/10.1016/j.future.2004.10.006}
{Future Generation Computer Systems \textbf{21} (2005) 549.}

\bibitem{kerberos} B.C.~Neumann and T.~Ts'o,
\textit{Kerberos: an authentication service for computer networks},
\href{http://dx.doi.org/10.1109/35.312841}
{IEEE Communications Magazine \textbf{32-9} (1994) 33}.

\bibitem{AFS} J.H.~Morris \etal,
\textit{Andrew: a distributed personal computing environment},
\href{http://dx.doi.org/10.1145/5666.5671}
{Commun.\ ACM \textbf{29-3} (1986) 184}.

\bibitem{GSI} I.~Foster \etal,
\textit{A Security Architecture for Computational Grids}
{Proc. 5th ACM Conference on Computer and Communications Security Conference, pp. 83-92, 1998.}


\bibitem{henderson_1995} R.L.~Henderson,
\textit{Job scheduling under the Portable Batch System},
\href{http://dx.doi.org/10.1007/3-540-60153-8}
{pp.~279-294 of: D.G.~Feitelson and L.~Rudolph (Eds.),
\textit{Job scheduling strategies for parallel processing}
[Lecture Notes in Computer Science \textbf{949}]} (Springer, Berlin, 1995).

\bibitem{schwickerath_2008} U.~Schwickerath and V.~Lefebure,
\textit{Usage of LSF for batch farms at CERN},
\href{http://dx.doi.org/10.1088/1742-6596/119/4/042025}
{J.~Phys.: Conf.\ Ser.\ \textbf{119} (2008) 042025}.

\bibitem{gentzsch_2001} W.~Gentzsch,
\textit{Sun Grid Engine: towards creating a compute power Grid},
\href{http://dx.doi.org/10.1109/CCGRID.2001.923173}
{pp.~35-36 of: R.~Buyya, G.~Mohay and P.~Roe (Eds.),
\textit{Proc.\ First IEEE/ACM International Symposium on Cluster
Computing and the Grid}} (IEEE Computer Society, Los Alamitos, CA, 2001).

\bibitem{thain_2005} D.~Thain, T.~Tannenbaum and M.~Livny,
\textit{Distributed computing in practice: the Condor experience},
\href{http://dx.doi.org/10.1002/cpe.938}
{Concurrency Computat.: Pract.\ Exper.\ \textbf{17} (2005) 323}.

\bibitem{ellert_2007} M.~Ellert \etal,
\textit{Advanced Resource Connector middleware for lightweight
computational Grids},
\href{http://dx.doi.org/10.1016/j.future.2006.05.008}
{Future Generation Computer Systems \textbf{23} (2007) 219}.
  
\bibitem{OSG} R.~Pordes \etal,
\textit{The Open Science Grid},
\href{http://dx.doi.org/10.1088/1742-6596/78/1/012057}
{J.~Phys.: Conf.\ Ser.\ \textbf{78} (2007) 012057}.

\bibitem{ROOT} R.~Brun and F.~Rademakers,
\textit{ROOT - an object oriented data analysis framework},
\href{http://dx.doi.org/10.1016/S0168-9002(97)00048-X}
{Nucl.\ Instrum.\ Methods \textbf{A389} (1997) 81}.

\bibitem{GoF}
  E.~Gamma \etal,
  \href{http://www.pearsonhighered.com/educator/academic/product/0,,0201633612,00%2Ben-USS_01DBC.html}
{\textit{Design patterns: elements of reusable object-orientated software}}
(Addison-Wesley, 1995).


\bibitem{JSDL}
\textit{Job Submission Description Language (JSDL) Specification, Version 1.0}
\href{http://www.gridforum.org/documents/GFD.56.pdf}{http://www.gridforum.org}

\bibitem{OGSA-BES}
\textit{OGSA Basic Execution Service}
\href{www.ogf.org/documents/GFD.108.pdf}{http://www.ogf.org}

\bibitem{SAGA}
\textit{A Simple API for Grid Applications (SAGA)}
\href{http://www.ogf.org/documents/GFD.90.pdf}{http://www.ogf.org}


\bibitem{IPython} F.~Perez and B.E.~Granger,
\textit{IPython: a system for interactive scientific computing},
\href{http://dx.doi.org/10.1109/MCSE.2007.53}
{Computing in Science and Engineering \textbf{9-3} (2007) 21}.

\bibitem{AMGA} B.~Koblitz, N.~Santos and V.~Pose,
\textit{The AMGA metadata service},
\href{http://dx.doi.org/10.1007/s10723-007-9084-6}
{J.~Grid Computing \textbf{6} (2008) 61}.

\bibitem{lhcb:2005jj}
R.~Antunes-Nobrega \etal\ [LHCb Collaboration],
\textit{LHCb computing},
\href{http://cdsweb.cern.ch/record/835156}
{Technical Design Report CERN/LHCC 2005-019 LHCb TDR-11 (2005).}

\bibitem{WebDav} E.J.~Whitehead Jr.,
\textit{World Wide Web Distributed Authoring and Versioning (WebDAV):
an introduction},
\href{http://dx.doi.org/10.1145/253452.253458}
{StandardView \textbf{5} (1997) 3.}

\bibitem{IMAP} P.~Heinlein and P.~Hartleban,
\href{http://nostarch.com/imap.htm}
{\textit{The book of IMAP: building a mail server with Courier and Cyrus}}
(No Startch Press, San Francisco, CA, 2008).

\bibitem{andreeva_2008} J.~Andreeva \etal,
\textit{Dashboard for the LHC experiments},
\href{http://dx.doi.org/10.1088/1742-6596/119/6/062008}
{J.\ Phys.\ Conf.\ Ser.\ \textbf{119} (2008) 062008}.

\bibitem{rempt_2001} B.~Rempt,
\href{http://www.commandprompt.com/community/pyqt/}
{\textit{GUI programming with Python: QT edition}}
(Command Prompt Inc, White Salmon, WA, 2001).

\bibitem{gaudi} G.~Barrand \etal,
\textit{GAUDI - a software architecture and framework for building HEP
data processing applications},
\href{http://dx.doi.org/10.1016/S0010-4655(01)00254-5}
{Computer Physics Communications \textbf{140} (2001) 45}.

\bibitem{DIRAC} A.~Tsaregorodtsev \etal,
\textit{DIRAC: a community grid solution},
\href{http://dx.doi.org/10.1088/1742-6596/119/6/062048}
{J.\ Phys.\ Conf.\ Ser.\ \textbf{119} (2008) 062048}.

\bibitem{DIRACAPI} S.~Paterson,
\textit{LHCb distributed data analysis on the computing Grid},
PhD Thesis, University of Glasglow (2006)
\href{http://cdsweb.cern.ch/record/995676/}
{[CERN-THESIS-2006-053]}.

\bibitem{RooFit} W.~Verkerke and D.~Kirkby,
\textit{The RooFit toolkit for data modeling},
\href{http://www.slac.stanford.edu/econf/C0303241/proc/papers/MOLT007.PDF}
{Contribution MOLT007 in:
Proc.\ 2003 Conference for Computing in High Energy and Nuclear Physics,
La Jolla, CA [SLAC eConf C0303241]}.

\bibitem{bib:atlascompmod} G.~Duckeck \etal~ (Eds.),
\textit{ATLAS computing},
\href{http://cdsweb.cern.ch/record/837738}
{Technical Design Report CERN/LHCC 2005-022 ATLAS TDR-017 (2005)}.

\bibitem{bib:atlasdq2} 
 M.~Branco \etal\,
\textit{Managing ATLAS data on a petabyte-scale with DQ2},
\href{http://dx.doi.org/10.1088/1742-6596/119/6/062017}
{J.~Phys.\ Conf.\ Ser.\ \textbf{119} (2008) 062017}.

\bibitem{maeno_2008} T.~Maeno,
\textit{PanDA: distributed production and distributed analysis system for
ATLAS},
\href{http://dx.doi.org/10.1088/1742-6596/119/6/062036}
{J.~Phys.\ Conf.\ Ser.\ \textbf{119} (2008) 062036}.

\bibitem{ballintijn_2006} M.~Ballintijn \etal,
\textit{Parallel interactive data analysis with PROOF},
\href{http://dx.doi.org/10.1016/j.nima.2005.11.100}
{Nucl.\ Instrum.\ Methods \textbf{559} (2006) 13}.

\bibitem{jones_2005} R.~Jones,
\textit{An overview of the EGEE project},
\href{http://dx.doi.org/10.1007/11549819}
{pp.~1-8 of: C.~T\"urker, M.~Agosti and H.-J.~Schek (Eds.),
\textit{Peer-to-peer, Grid, and service-orientation in digital library
architectures}
[Lecture Notes in Computer Science \textbf{3664}]} (Springer, Berlin, 2005).

\bibitem{ellert_2003} M.~Ellert \etal,
\textit{The NorduGrid project: using Globus toolkit for building Grid
infrastructure}
\href{http://dx.doi.org/10.1016/S0168-9002(03)00453-4}
{Nucl.\ Instrum.\ Methods \bf{A502} (2003) 407}.

\bibitem{town_2004} C.~Town and D.~Sinclair,
\textit{Language-based querying of image collections on the basis of an
extensible ontology},
\href{http://dx.doi.org/10.1016/j.imavis.2003.10.002}
{Image and Vision Computing \textbf{22} (2004) 251}.

\bibitem{Garfield} R.~Veenhof,
\textit{Garfield - simulation of gaseous detectors},
\href{http://consult.cern.ch/writeup/garfield/}
{CERN Program Library User Guide W5050} (1984 \textit{et seq.}).

\bibitem{DIANE} J.T.~Mo\'scicki,
\textit{Distributed analysis environment for HEP and interdisciplinary
applications},
\href{http://dx.doi.org/10.1016/S0168-9002(03)00459-5}
{Nucl.\ Instrum.\ Methods \textbf{A502} (2003) 426}.

\bibitem{Geant4MedicalDIANE} J.T.~Mo\'scicki \etal,
\textit{Distributed Geant4 Simulation in Medical and Space Science Applications
using DIANE framework and the GRID},
{Nucl.\ Phys. B (Proc. Suppl.) \textbf{125} (2003) 327-331}

\bibitem{Geant4} J~.Allison \etal,
\textit{Geant 4 - a simulation toolkit},
\href{http://dx.doi.org/10.1016/S0168-9002(03)01368-8}
{Nucl.\ Instrum.\ Methods \textbf{A506} (2003) 250}.

\bibitem{AvianFlu} H.-C.\ Lee \etal,
\textit{Grid-enabled high-throughput in silico screening against influenza A
neuraminidase},
\href{http://dx.doi.org/10.1109/TNB.2006.887943}
{IEEE Trans.\ NanoBioscience \textbf{5-4} (2006) 288}.

\bibitem{ITU} J.T.~Mo\'scicki \etal,
\textit{ITU RRC06 on the Grid},
{in prep. for Journal of Grid Computing }

\bibitem{LQCD} J.T.~Mo\'scicki \etal,
\textit{Lattice QCD on the Grid}
{in prep. for Computer Physics Communications}

\bibitem{SchedulingForResponsiveGrids}
  C.~Germain-Renaud, C.~Loomis , J.~T.~Mo{\'s}cicki and R.~Texier,
\textit{Scheduling for Responsive Grids}, 
\href{http://dx/doi.org/10.1007/s10723-007-9086-4}
{J. Grid Computing \textbf{6}, (2008) 15-27 }

\bibitem{GridWay}
J.~Herrera \etal,
\textit{Porting of Scientific Applications to Grid Computing on GridWay}
{Scientific Programming \textbf{13} 4 , pp. 317-331, 2005}

\bibitem{MCS}
R.P.~Bruin \etal,
\textit{Job submission to grid computing environments}
\href{httpL//dx/doi.org/10.1002/cpe.1290}
{Concurrency and Computation: Pract. and Exper. \textbf{20} 11, pp. 1329-1340 (2008)}






\end{thebibliography}
\end{document}